\documentstyle[pre,aps,epsf,float,multicol]{revtex}
\begin{document}
\draft
\title{\bf Non-universal size dependence of the free energy of 
confined systems near criticality}
\author{X.S. Chen$^{1,2,3}$ and V. Dohm$^{2}$}
\address{$^1$ Institute of Theoretical Physics, Academia Sinica,
P.O. Box 2735, Beijing 100080, China}
\address{$^2$ Institut f\"{u}r Theoretische Physik, Technische Hochschule
Aachen, D-52056 Aachen, Germany}
\address{$^3$ Institut f\"ur Festk\"orperforschung,
Forschungszentrum J\"ulich, D-52425 J\"ulich, Germany}
\date{\it \today}
\maketitle
\begin{abstract}

The singular part of the finite-size free energy density $f_s$
of the O$(n)$ symmetric $\varphi^4$ field theory in the large-$n$ limit
is calculated at finite cutoff for confined geometries of linear size 
$L$ with periodic boundary conditions in $2 < d < 4$ dimensions. 
We find that a sharp cutoff $\Lambda$ causes a non-universal leading 
size dependence $f_s \sim \Lambda^{d-2} L^{-2}$ near $T_c$ which dominates
the universal scaling term $\sim L^{-d}$. This implies a non-universal 
critical Casimir effect at $T_c$ and a leading non-scaling term 
$\sim L^{-2}$ of the finite-size specific heat above $T_c$.

\end{abstract}
\pacs{PACS numbers: 05.70.Jk, 64.60.-i}
\begin{multicols}{2}
The concept of universal finite-size scaling has played an important role in
the investigation of finite-size effects near critical points over
the last decades \cite{fisher,barber,finite}. Consider the
free-energy density $f (t, L)$ of a finite system at the reduced
temperature $t = (T - T_c) / T_c$ and at vanishing external field
in a $d$-dimensional cubic geometry of volume $L^d$ with periodic
boundary conditions. It is well known that, for small $t$, the bulk
free energy density $f_b \equiv f (t, \infty)$ can be decomposed as
\begin{equation}
\label{gleichung0} f_b (t) = f_{bs} (t) + f_0 (t)
\end{equation}
where $f_{bs} = A^\pm | t |^{2-\alpha}$ denotes the singular part of
$f_b$ above $(+)$ and below $(-)$ $T_c$, apart from bulk
corrections to scaling, and where the regular part $f_0 (t)$ of $f_b$
can be identified unambiguously. 

According to Privman and Fisher
\cite{privman-fisher,privman} the singular part of the finite-size
free-energy density may be defined by
\begin{equation}
\label{gleichung1} f_s (t, L) = f (t, L) - f_0 (t)
\end{equation}
where $f_0$ is independent of $L$. The finite-size scaling
hypothesis asserts that, below the upper critical dimension $d =
4$ and in the absence of long-range interactions, $f_s (t, L)$
has the structure \cite{privman-fisher,privman,see also}
\begin{equation}
\label{gleichung2} f_s (t, L) = L^{-d} \; {\cal F} (L/\xi)
\end{equation}
where ${\cal F}(x)$ is a universal scaling function and $\xi = \xi_0^
\pm |t|^{-\nu}$ is the bulk correlation length. Both $\xi$
and $L$ are assumed to be sufficiently large compared to
microscopic lengths (for example, the lattice spacing $\tilde a$ of
lattice models, the inverse cutoff $\Lambda^{-1}$ of field theories,
or the length scale of subleading long-range interactions). 
Eqs.(\ref{gleichung0}) - (\ref{gleichung2}) are expected to remain
valid also for non-cubic geometries where the scaling function
${\cal F} (x)$ depends on the geometry and on the universality
class of the bulk critical point but not on $\tilde a$ or $\Lambda$
and not on other interaction details 
\cite{privman-fisher,privman,see also,privman1}.

As a consequence, universal finite-size scaling properties are generally 
believed to
hold for quantities derived from $f_s$, such as the critical
Casimir force in film geometry \cite{fisher-gennes,krech,brankov} and the
finite-size specific heat $C_s$, in apparent agreement with 
renormalization-group
and model calculations \cite{krech-dietrich,esser,danchev,borjan}. In
the field-theoretic calculations \cite{krech-dietrich,esser}, however,
the method of dimensional regularization was employed and no proof was 
given for the unimportance of the cutoff dependence of
$f_s$ and $C_s$.

In this Letter we reexamine the validity of the scaling prediction
Eq.(\ref{gleichung2}) on the basis of exact results for the O$(n)$ symmetric
$\varphi^4$ field theory {\it {at finite cutoff}} in the large-$n$ limit.
We shall show that $f_s$ depends significantly on the cutoff procedure : 
a sharp cutoff
$\Lambda$ in ${\bf k}$ space causes non-negligible finite-size effects
above and at $T_c$ that violate the universal scaling form 
Eq.(\ref{gleichung2}).
Specifically, we find that Eq.(\ref{gleichung2}) must be complemented as
\begin{equation}
\label{gleichung3} f_s (t, L, \Lambda) \; = \; L^{-2} \; \Lambda^{d-2} \; \Phi
(\xi^{-1} \Lambda^{-1}) \; + \; L^{-d} \; {\cal F} (L / \xi)
\end{equation}
where the function $\Phi$ has a finite geometry-dependent critical value
$\Phi (0) > 0$. For $d > 2$, the non-scaling $L^{-2}$ term exhibits a 
dominant size
dependence compared to the $L^{-d}$ scaling term. By contrast, for the
$\varphi^4$ {\it lattice} model with short-range interaction, we find
that no $L^{-2}$ term exists for $\hat{f}_s (t, L, \tilde a)$ and that 
Eq.(\ref{gleichung2}) is indeed valid except that for $L \gg \xi$ the 
exponential
scaling argument \cite{see also} of ${\cal F}$  must be formulated 
in terms of the lattice-dependent
"exponential" correlation length \cite{chen-dohm-2000,fisher-burford}.

Non-negligible cutoff and lattice effects were already found previously
for the finite-size susceptibility \cite{chen-dohm-1999,chen-dohm-99}.
Similar non-universal finite-size effects were found in systems
with subleading long-range interactions \cite{dantchev}. These
effects, however, were restricted to the regime $L \gg \xi$ close
to the bulk limit above $T_c$. The new non-scaling finite-size effect
exhibited in Eq.(\ref{gleichung3}) is significantly more general in
that it is pertinent to the entire $\xi^{-1} - L^{-1}$ plane including
the central finite-size regime $\xi \gg L$. In particular, this effect
exists at $T_c$ where
\begin{equation}
\label{gleichung4} f_s (0, L, \Lambda) = L^{-2} \; \Lambda^{d-2} \; \Phi (0)
\; + \; L^{-d} \; {\cal F} (0)
\end{equation}
has a non-universal leading amplitude $\Lambda^{d-2} \; \Phi
(0)$ for $d > 2$. As an important consequence, this implies the
non-universality of the critical Casimir force $L^{-2}
\Lambda^{d-2} \; \Phi_{film} (0)$ in film geometry. Furthermore
this indicates that material-dependent properties such as subleading
long-range interactions \cite{dantchev} (which exist in real systems
and which do not change the universal {\it bulk} properties) may affect
the leading $L$ dependence of $f_s$.

We start from the standard $\varphi^4$ continuum Hamiltonian
\begin{equation}
\label{gleichung5} H = \int\limits_V d^d x
\left[\frac{1}{2}\; r_0 \; \varphi^2 \; + \; \frac{1}{2} \; (\nabla
\varphi)^2 + u_0 (\varphi^2)^2 \right]
\end{equation}
with $r_0 = r_{0c} + a_0 t$ for the $n$-component field $\varphi
({\bf x})$ in the cubic volume $V = L^d$ with periodic boundary
conditions. This model requires a specification of the ${\bf x}$
dependence of $\varphi ({\bf x})$ at short distances.
As usual, we assume that the Fourier amplitudes
$\hat \varphi_{\bf k} = \int_V d^d x \; \varphi ({\bf x}) e^{- i
{\bf kx}}$ are restricted to wave vectors $\bf k$ with components
$k_j$ in the range $- \Lambda \leq k_j < \Lambda$ where $k_j = 2
\pi m_j / L$, $j = 1,2,...,d$, with $m_j = 0, \pm 1, \pm 2, ...$ .
The question can be raised whether there exists a non-negligible
cutoff dependence of the finite-size free energy density per component
(divided by $k_B T)$,
\begin{equation}
\label{gleichung5a} f_{cube} (t, L, \Lambda) = - n^{-1} L^{-d} \; \ln
Z_{cube}
\end{equation}
where
\begin{equation}
\label{gleichung6} Z_{cube} (t, L, \Lambda) = \prod_{\bf k} \; \int\limits
\frac{d  \;\hat \varphi_{\bf k}}{\Lambda^{(d-2)/2}} \; \exp (- H)
\end{equation}
is the dimensionless partition function. For comparison we shall
also consider the free energy density $\hat f (t, L, \tilde a)$ of
the $\varphi^4$ lattice model
\begin{equation}
\label{gleichung7} \hat H = \tilde a^d \left[\sum_i \left(\frac{r_0}{2}
\varphi^2_i + u_0 (\varphi_i^2)^2 \right) + \sum_{< i j >} 
\frac{J}{2 \tilde a^2} 
(\varphi_i - \varphi_j)^2 \right]
\end{equation}
with a nearest-neighbor coupling $J$ on a simple-cubic lattice with
a lattice spacing $\tilde a$. The factor $(k_B T)^{-1}$ is absorbed
in $H$ and $\hat H$.

Essential features of the cutoff effects on the finite-size free
energy can already be demonstrated within the simple Gaussian
model Eq.(\ref{gleichung5}) with $u_0 = 0$ which implies $r_{0c} = 0$ 
and $\xi = r_0^{-1/2}$. Straightforward integration leads to
\begin{equation}
\label{gleichung8} f_{cube}^{(G)} = \frac{1}{2} \left\{L^{-d} \sum_{\bf k} 
\ln [\Lambda^{-2} (r_0 + k^2)] - \Lambda^d
\ln \pi - L^{-d} \ln 2 \right\} .
\end{equation}
In the bulk limit this yields
\begin{equation}
\label{gleichung9} f_b^{(G)} \; = \; - \; \frac{1}{2} \; \Lambda^d
\ln \pi \; + \;
\frac{1}{2} \int\limits_{\bf k} \ln \left[\Lambda^{-2} \left(r_0 + k^2
\right)\right]
\end{equation}
where $\int_{\bf k}$ stands for $(2 \pi)^{-d} \int d^d k$ with $|k_j|
\leq \Lambda$. The regular part reads $f_0^{(G)} \; = \; (\tilde c_1 +
\tilde c_2 \; r_0) \Lambda^d$ where $\tilde c_1$ and $\tilde c_2$ are
$d$ dependent constants.

In $2 < d < 4$ dimensions, the finite sharp cutoff $\Lambda$ causes only
negligible corrections to the (leading) singular part
$f_{bs}^{(G)} = Y^{(G)} \; \xi^{-d}$ of the {\it bulk} free energy
density where $Y^{(G)}$ is a universal amplitude . For the {\it finite}
system, however, the $\Lambda$ dependence turns out to be non-negligible.
For large $\xi \Lambda$ and $L \Lambda$ at fixed $\Lambda$ we find that
the finite-size free energy $f_s^{(G)} = f^{(G)} - f_0^{(G)}$ attains the
form of Eq.(\ref{gleichung3}) with the non-scaling contribution
\begin{eqnarray}
\label{gleichung14} \Phi_{cube} (\xi^{-1} \Lambda^{-1}) &=&
\frac{d}{6 (2 \pi)^{d-2}} \int\limits^\infty_0 d y
\left[\int\limits^1_{-1} d q \; e^{- q^2 y} \right]^{d-1}\times \nonumber \\
&\times&\exp \left[- (1 + \xi^{-2} \Lambda^{-2}) y \right]
\end{eqnarray}
and with the universal scaling function
\begin{eqnarray}
\label{gleichung13} {\cal F}_{cube}^{(G)} (L/\xi) &=& - \frac{\ln 2}{2} 
+ (L/\xi)^d \; Y^{(G)} \nonumber \\
&&+ \frac{1}{2}
\int\limits^\infty_0 \frac{d y}{y} \; W_d (y) \; e^{- (L/\xi)^2 y / 4
\pi^2}, \\
\label{gleichung13a} W_d (y) &=& \left(\frac{\pi}{y} \right)^{d/2}
\; - \; \left(\sum^\infty_{m = - \infty} e^{- y m^2}\right)^d , \\
\label{gleichung10} Y^{(G)} &=& - \frac{A_d}{d (4-d)} \; ,
\;\;\; A_d \; = \; \frac{\Gamma (3 - d/2)}{2^{d-2} \;
\pi^{d/2} (d-2)} \; .
\end{eqnarray}
We note that ${\cal F}_{cube}^{(G)} (L/\xi)$ diverges in the critical limit
$L/\xi \rightarrow 0$ which is an artifact of the Gaussian model for
cubic geometry (arising from the isolated ${\bf k} = {\bf 0}$ term in
$f^{(G)}_{cube}$ ).

Here we shall show that the non-scaling structure of 
Eqs.(\ref{gleichung3}) and (\ref{gleichung14}) is valid more
generally, beyond the Gaussian model, for the $\varphi^4$ field
theory. In the limit $n \rightarrow \infty$ at fixed $u_0 n$, the
free energy density is \cite{chen-dohm-529}
\begin{eqnarray}
\label{gleichung15} f_{cube} (t, L, \Lambda) = &-& 
\frac{\ln \pi}{2} \Lambda^d   \; -\frac{\ln 2}{2} L^{-d}   \;
 - \frac{(r_0 - \chi^{-1})^2}
{16 u_0 n} \nonumber \\
&+& \frac{1}{2} \; L^{-d} \sum_{\bf k} \ln \left[\Lambda^{-2}(\chi^{-1}
+ {\bf k}^2 )\right]
\end{eqnarray}
where $\chi^{-1}$ is determined implicitly by
\begin{equation}
\label{gleichung16} \chi^{-1} = r_0 + 4 u_0 n \; L^{-d} \sum_{\bf k}
\left(\chi^{-1} + {\bf k}^2 \right)^{-1} \; .
\end{equation}
The bulk free energy $f_b$ and bulk susceptibility $\chi_b$ are obtained
by the replacement $L^{-d} \sum_{\bf k} \rightarrow \int_{\bf k}$,
and the critical point is determined by $r_0 = r_{0c} = - 4 u_0 n
\int_{\bf k} {\bf k}^{-2}\; $. The regular part of $f_b$ reads $f_0 \; = \;
\tilde c_1 \; \Lambda^d \; - \; r_0^2 / (16 u_0 n)$.

For the singular part $f_s = f - f_0$ of the finite-size free
energy above and at $T_c$ we again find the form of Eq.(\ref{gleichung3}).
The leading non-scaling part $\Phi_{cube} (\xi^{-1} \Lambda^{-1})$ turns
out to have the same form as that of the Gaussian model, 
Eq.(\ref{gleichung14}), except that now, in the large-$n$ limit,
the bulk correlation length above $T_c$ is
\begin{equation}
\label{gleichung16a} \xi = \chi_b^{1/2} = \xi_0 t^{- \nu}, \; \nu =
(d-2)^{-1} 
\end{equation}
with $\xi_0 = \left\{ 4u_0 n A_d / [a_0 (4-d)] \right\}^{1/(d-2)}$.

For the subleading universal scaling part we find for $2 < d < 4$
\begin{eqnarray}
\label{gleichung19} {\cal F}_{cube} (L/\xi) = &-& \frac{\ln 2 }{2} +
\frac{A_d}{2 (4-d)} \left[(L/\xi)^{d-2} P^2 - \frac{2}{d} \; P^d
\right] \nonumber\\[12pt] 
&+& \frac{1}{2} \int\limits^\infty_0 \frac{d y}{ y }
W_d(y) \; e^{- P^2 y / 4 \pi^2}
\end{eqnarray}
where $P(L/\xi)$ is determined implicitly by
\begin{equation}
\label{gleichung19a} P^{d-2} = (L/\xi)^{d-2} - 
\frac{4-d}{4 \pi^2 A_d} \int\limits^\infty_0 d y \; W_d (y) \; e^{-
P^2 y / 4 \pi^2} .
\end{equation}
In the bulk limit Eqs.(\ref{gleichung19}) and (\ref{gleichung19a})
yield $f_{bs} = Y \xi^{-d}$ with 
\begin{equation}
\label{gleichungY} Y = (d-2) A_d / [2d(4-d)].
\end{equation}

These results can be extended to partially confined 
$L^{d'} \times \infty^{d-d'}$
geometries with periodic boundary conditions in
$d' < d$ dimensions.  Eq.(\ref{gleichung3}) remains valid where now the 
non-scaling part reads for the Gaussian model
\begin{equation}
\label{gleichung29} \Phi_{d,d'} (\xi^{-1} \Lambda^{-1}) =
\frac{d'}{d} \; \Phi_{cube} (\xi^{-1} \Lambda^{-1})
\end{equation}
with $\Phi_{cube} (\xi^{-1} \Lambda^{-1})$ given by
Eq.(\ref{gleichung14}). For the finite-size scaling function of the
Gaussian model we find
\begin{eqnarray}
\label{gleichung28} {\cal F}^{(G)}_{d,d'} (L/\xi) = (L/\xi)^d
\; Y^{(G)} &+& \frac{1}{2} \int\limits_0^\infty \frac{d y}{ y }
\left(\frac{\pi}{y}\right)^{(d-d')/2} \times \nonumber \\
&\times& \; W_{d'} (y) e^{- (L/\xi)^2 y/ 4 \pi^2}.
\end{eqnarray}
For the $\varphi^4$ theory in the large-$n$ limit the corresponding result is
in $2 < d < 4$ dimensions
\begin{eqnarray}
\label{gleichung30} {\cal F}_{d,d'} (L/\xi) &=& \;
\frac{A_d}{2 (4-d)} \left[(L/\xi)^{d-2} P^2 - \frac{2}{d} \; P^d
\right] \nonumber\\[12pt] &+& \frac{1}{2} \int\limits^\infty_0 
\frac{d y}{ y } \left(\frac{\pi}{y}\right)^{(d-d')/2} \;
W_{d'}(y) \; e^{- P^2 y / 4 \pi^2}
\end{eqnarray}
where now $P (L/\xi)$ is determined implicitly by
\begin{eqnarray}
\label{gleichung31} P^{d-2} \; =\; (L/\xi)^{d-2} 
\; &-& \;\frac{4-d}{4 \pi^2 A_d} \int\limits^\infty_0 d y \;
\left(\frac{\pi}{y}\right)^{(d-d')/2} \times \nonumber \\
&&\times \; W_{d'} (y) e^{- P^2 y / 4 \pi^2} \; .
\end{eqnarray}
We find that the non-scaling part $\Phi_{d,d'}(\xi^{-1} \Lambda^{-1})$ in 
the large-$n$ limit has the same form as the Gaussian result, 
Eq.(\ref{gleichung29}), with $\xi$ given by Eq.(\ref{gleichung16a}).

These results have a significant consequence for the critical
Casimir effect. In film geometry $(d' = 1)$ the Casimir force
is defined as \cite{krech,brankov}
\begin{equation}
\label{gleichung31-1} F_{Casimir} = - \partial f^{ex}_{film} (t, L,\Lambda)
/ \partial L
\end{equation}
where the excess free energy per unit area is given by
\begin{equation}
\label{gleichung31-2} f^{ex}_{film} (t, L, \Lambda) = L f_{film}
(t, L, \Lambda) - L f_b (t)\;.
\end{equation}
Near bulk criticality our results, Eqs.(\ref{gleichung29}),
 (\ref{gleichung30}) and (\ref{gleichung31}), yield for $d' = 1$
\begin{eqnarray}
\label{gleichung31a} F_{Casimir} (L, \xi, \Lambda) &=& L^{-2}
\Lambda^{d-2} \Phi_{d,1} (\xi^{-1}\Lambda^{-1}) \nonumber \\
&& + \; L^{-d} \; X_{Casimir}(L/\xi)
\end{eqnarray}
where
\begin{equation}
\label{gleichung31aa} X_{Casimir}(x) = (d-1) {\cal F}_{d,1}(x) 
-  x \; {\cal F}'_{d,1} (x) + Y x^{d}
\end{equation}
with ${\cal F}' (x) = \partial {\cal F} (x) / \partial x$.
Thus the critical Casimir force  has a {\it leading} non-universal 
term $\sim L^{-2}$, in
addition to the {\it subleading} universal terms $\sim L^{-d}$ 
of previous theories \cite{krech,brankov,krech-dietrich,danchev}.

We have also calculated the cutoff dependence of the finite-size
specific heat $C (t, L, \Lambda)$ above $T_c$. For example 
for cubic geometry and in the large-$n$ limit we find, at fixed 
$0 < t \ll 1$, the large-$L$-behavior
\begin{equation}
\label{gleichung31b} C (t, L, \Lambda) \; - \; C_b (t) = 
L^{-2} \Lambda^{d-2} \Psi (\xi \Lambda)+ O(e^{- L/\xi})
\end{equation}
where
\begin{equation}
\label{gleichung31c} \Psi (\xi\Lambda) = B_0 \;(\xi \Lambda)^{2(d-3)} 
\Phi_{cube} (\xi^{-1}\Lambda^{-1})
\end{equation}
with $B_0 = (4-d)\; (\xi_0\Lambda)^{2(2-d)}/(d-2)^2$.
A similar result holds for $L^{d'} \times \infty^{d-d'}$ geometries.

Eqs.(\ref{gleichung3}), (\ref{gleichung14}) and (\ref{gleichung29}) -
(\ref{gleichung31c}) are the main results of the present paper.
We have confirmed the structure of these results  also for
the $\varphi^4$ theory with {\it finite} $n$ within a one-loop
renormalization-group calculation at finite $\Lambda$ which yields 
the same form of the function $\Phi_{d,d'} (\xi^{-1} \Lambda^{-1})$ 
as in Eq.(\ref{gleichung29}).
Thus we arrive at the general conclusion that there exists no universal 
finite-size scaling form for the {\it leading} size dependence of the 
free energy density $f_s$ near criticality of confined systems with 
periodic boundary conditions, contrary to phenomenological expectations 
\cite{privman-fisher,privman,see also}. Clearly this raises the necessity
of reexamining also the universality predictions for the leading 
finite-size effects in systems with {\it non-periodic} boundary conditions
\cite{fisher,barber,finite,privman-fisher,privman,krech-dietrich,vdohm}.

The sensitivity of $f_s (t, L, \Lambda)$ with respect to the cutoff
procedure can be related to a corresponding sensitivity of the
{\it bulk} correlation function $G ({\bf x}) = \; < \varphi ({\bf x})
\varphi (0) >$ in the range $|{\bf x}| \gg \xi$ \cite{chen-dohm-2000}.
For example, for the $\varphi^4$ continuum Hamiltonian Eq.(\ref{gleichung5}) 
with an isotropic sharp cutoff $|{\bf k}| \leq \Lambda$ we find, 
in the large-$n$ limit, the oscillatory power-law decay above $T_c$
\begin{eqnarray}
\label{gleichung24} G ({\bf x}) &=& 2 \Lambda^{d-2} 
(2 \pi x \Lambda)^{- (d+1)/2}
\;  \; \frac{\sin \; [\Lambda x - \pi (d-1) /
4]}{1 + \xi^{-2} \Lambda^{-2}} \nonumber \\
&& \; + \; O \left(e^{- x / \xi}\right)
\end{eqnarray}
for large $x = |{\bf x}| \gg \xi$ corresponding to the existence of
long-range spatial correlations. (A different power-law decay
is obtained for an anisotropic sharp cutoff $|k_j|\leq \Lambda$.)
By contrast, $G ({\bf x})$ has an exponential
decay for the lattice model Eq.(\ref{gleichung7}) with purely
short-range interaction \cite{chen-dohm-2000}. For the latter case we
find that the scaling form Eq.(\ref{gleichung2}) is valid provided that the
second-moment bulk correlation length $\xi$ in the argument of
${\cal F}$ is replaced by the lattice-dependent exponential
correlation length \cite{chen-dohm-2000,fisher-burford}. Specifically
we find for the Gaussian lattice Hamiltonian Eq.(\ref{gleichung7}),
at fixed $t > 0$, the exponential large-$L$ behavior $\hat{f}_s 
(t, L, \tilde a) = L^{-d} \; \hat{{\cal F}}_{cube} (L/\xi_1)$ with
\begin{equation}
\label{gleichung25} \hat{{\cal F}}_{cube} (L/\xi_1) = d (L / 2 \pi
\xi_1)^{(d-1)/2} \; \exp (- L / \xi_1)
\end{equation}
where $\xi_1 = (\tilde a / 2) \;  [{\rm arcsinh} (\tilde a / 2 \xi)
]^{-1}$ is the exponential correlation length in the direction of
one of the cubic axes. A similar result holds in the large-$n$
limit. Note that the {\it non-universal} dependence of $\xi_1$ on
$\tilde a$ is non-negligible in the exponent of Eq.(\ref{gleichung25})
\cite{chen-dohm-2000}.

As pointed out by Dantchev and Rudnick \cite{dantchev}, the
presence of subleading long-range interactions causes leading
non-scaling finite-size effects on the susceptibility similar
to those for the case of a sharp cutoff in the regime $L \gg \xi$
\cite{chen-dohm-1999,chen-dohm-99}. Clearly it would be desirable
to extend our present analysis to the case of such interactions.
On the basis of our present results we expect the existence of
leading non-scaling finite-size effects on $f_s$, for both continuum and
lattice systems with subleading long-range interactions, not only
for $L \gg \xi$ but also in the central finite-size regime
$\xi \gg L$ including $T = T_c$. In particular we expect a leading
non-universal size dependence of the critical Casimir effect in
such systems, different from the $L^{-d}$ scaling prediction.

Furthermore, according to Eq.(\ref{gleichung31b}), there exists a
non-negligible non-universal (power-law) finite-size contribution
$\sim (\Lambda L)^{-2}$ also for the singular part of the finite-size
specific heat in the region $L \gg \xi$ where the scaling contribution
$\sim e^{- L/\xi}$ is exponentially small for the case of periodic
boundary conditions. Corresponding non-scaling terms due to subleading
long-range interactions are potentially important  for the analysis of
the leading and subleading $L$ dependence of finite-size effects in
real systems with non-periodic boundary conditions.

Support by DLR and by NASA under contract numbers 50WM9911 and
1226553 is acknowledged.

\end{multicols}

\begin{references}
%
\bibitem{fisher}
M.  E. Fisher, in {\it Critical Phenomena, Proceedings of the 1970
International School of Physics ``Enrico Fermi'',}  Course 51,
edited by M. S. Green (Academic, New York, 1971), p. 1.
%
\bibitem{barber}
M. N. Barber, in {\it Phase Transitions and Critical Phenomena}, edited
by C. Domb, J.L. Lebowitz (Academic, New York, 1983),
Vol.~8, p. 145.
%
\bibitem{finite}
{\it Finite Size Scaling and Numerical Simulation of Statistical
Systems}, edited by V. Privman (World Scientific, Singapore, 1990).
%
\bibitem{privman-fisher}
V. Privman and M.E. Fisher, Phys. Rev. B {\bf 30}, 322 (1984).
%
\bibitem{privman}
V. Privman, in Ref. [3].
%
\bibitem{see also}
See also Sec. 1.2 of M.P. Gelfand and M.E. Fisher, Physica A {\bf
166}, 1 (1990).
%
\bibitem{privman1}
V. Privman, A. Aharony, and P.C. Hohenberg, in {\it Phase Transitions
and Critical Phenomena}, edited by C. Domb and J.L. Lebowitz
(Academic, New York, 1991), Vol. 14, p. 1.
%
\bibitem{fisher-gennes}
M.E. Fisher and P.G. de Gennes, C.R. Acad. Sci. Paris, Ser. B
{\bf 287}, 207 (1978).
%
\bibitem{krech}
M. Krech, {\it The Casimir Effect in Critical Systems} (World
Scientific, Singapore, 1994); J. Phys. Cond. Mat. {\bf 11}, R 391
(1999).
%
\bibitem{brankov}
J.G. Brankov, D.M. Danchev, and N.S. Tonchev, {\it The Theory of
Critical Phenomena in Finite-Size Systems - Scaling and Quantum
Effects} (World Scientific, Singapore, 2000).
%
\bibitem{krech-dietrich}
M. Krech and S. Dietrich, Phys. Rev. Lett. {\bf 66}, 345 (1991);
Phys. Rev. A {\bf 46}, 1886 (1992); {\bf
46}, 1922 (1992).
%
\bibitem{esser}
A. Esser, V. Dohm, and X.S. Chen, Physica A {\bf 222}, 355 (1995).
%
\bibitem{danchev}
D. M. Danchev, Phys. Rev. E {\bf 53}, 2104 (1996); E {\bf 58},
1455 (1998).
%
\bibitem{borjan}
Z. Borjan and P.J. Upton, Phys. Rev. Lett. {\bf 81}, 4911 (1998).
%
\bibitem{chen-dohm-2000}
X.S. Chen and V. Dohm, Eur. Phys. J. B {\bf 15}, 283 (2000).
%
\bibitem{fisher-burford}
M.E. Fisher and R.J. Burford, Phys. Rev. {\bf 156}, 583 (1967).
%
\bibitem{chen-dohm-1999}
X.S. Chen and V. Dohm, Eur. Phys. J. B {\bf 7}, 183 (1999).
%
\bibitem{chen-dohm-99}
X.S. Chen and V. Dohm, Eur. Phys. J. B {\bf 10}, 687 (1999).
%
\bibitem{dantchev}
D. Dantchev and J. Rudnick, Eur. Phys. J. B {\bf 21}, 251 (2001).
%
\bibitem{chen-dohm-529}
X.S. Chen and V. Dohm, Eur. Phys. J. B {\bf 5}, 529 (1998)
%
\bibitem{vdohm}
V. Dohm, Phys. Scr.  {\bf T49}, 46 (1993).
%
\end{references}
\end{document}